Roman GIELERAK and Jacek DAMEK[*]

# STABILITY OF THE BOSE-EINSTEIN CONDENSATE UNDER POLYNOMIAL PERTURBATIONS

*Abstract*

*The problem of the Bose-Einstein condensate preservation under thermofield and standard gauge-invariant perturbations is discussed. In particular, a new result on the stability of Bose-Einstein condensate under thermofield perturbations of a polynomial type is presented. The approach adopted here strongly relies on previous papers, where functional integral methods were employed in order to study the case of bounded thermofield-like perturbations. Additionally, a description of noncritical Bose matter originating from the ideal Bose gas after switching on the standard, two-body gauge-invariant interactions is presented in terms of Gibbsian-like perturbations of the Poissone-Wiener background measure.*

## I. INTRODUCTION: THE PROBLEM OF THE BOSE – EINSTEIN CONDENSATE

The problem of the Bose-Einstein condensate preservation that does arise when one tries to analyze the free Bose Matter under switching on any realistic two-body interaction between particles has still remained open despite of its long history and efforts of many theorists. Even in the case of noninteracting Bose particles, the process of providing a complete and rigorous (from the mathematical physics point of view) proof of Einstein's heuristic indications [1] took a very long time during which Mark Kac's fundamental contribution [2] played a crucial role. His ideas have been cleaned up due to the long-time activity of the Dublin group and, at present, a very clear, general understanding of the condensation phenomenon in noninteracting Bose systems is available [3].

Let $(\Lambda_n)$ be a monotonous family of bounded regions in the Euclidean space $R^d, d \geq 3$, such that $\bigcup_n \Lambda_n = R^d$ and let $(h_n)$ be a family of one-particle kinetic energy operators that possess purely discrete spectra

---

[*] Institute of Mathematics Technical University of Zielona Góra Poland
The authors supported by the Grant of the Rector of Technical University of Zielona Góra

$$\sigma(h_n) = \{\lambda_n(k)\}_k$$

with the lowest eigenvalues $\lambda_n(1) \geq 0$ for all $n$. We also define

$$\sigma_n(k) = \lambda_n(k) - \lambda_n(1), \text{ for } k = 1, 2, \ldots$$

Then we will say that a family $(\Lambda_n, h_n)$ is admissible if

$$\begin{cases} \text{for all } \beta > 0, \lim_n \frac{1}{|\Lambda_n|} \sum_{k=1}^{\infty} \exp(-\beta \sigma_n(k)) \equiv \phi^{\infty}(\beta) \\ \text{exists and } \phi^{\infty}(\beta) \neq 0 \text{ for some } \beta \in (0, \infty). \end{cases} \quad (1)$$

The following result (formulated informally) seems to be the most general one for noninteracting systems [3]. Let us define the partition function $Z_n(\beta, \mu)$ by

$$Z_n(\beta, \mu) = \text{Tr}_{\Gamma(L^2(\Lambda_n, dx))} \exp(-\beta \, d\Gamma(h_n + \mu 1)), \quad (2)$$

and the finite volume free energy density:

$$p_n(\beta, \mu) = \frac{1}{|\Lambda_n|} \ln Z_n(\beta, \mu) \quad (3)$$

where $|\Lambda|$ stands for the volume of $\Lambda$ and $\Gamma(L^2(\Lambda, dx))$ is the bosonic Fock space on $L^2(\Lambda, dx)$.

**Theorem** ([3]). *Let $(\Lambda_n, h_n)$ be an admissible family. Then*
(1) *the unique limit*

$$p_{\infty}(\beta, \mu) = \lim_n p_n(\beta, \mu)$$

*exists, and*
(2) *the equation*

$$\frac{d}{d\mu} p_{\infty}(\beta, \mu) = \rho$$

*has a unique solution $\mu$ for any $\rho \in (o, \infty)$, and such that for any $\rho \geq \rho_{cr}(\beta)$,*

*where*

$$\rho_{cr}(\beta) = \int_{(0,\infty)} (e^{\beta\lambda} - 1)^{-1} dF(\lambda), \qquad (4)$$

*the corresponding µ equals to zero.*
*The density of states dF in (4) is uniquely determined by*

$$\phi^{\infty}(\beta) = \int_{(0,\infty)} e^{-\beta\lambda} dF(\lambda) \qquad (5)$$

Now, let $\Lambda_n$ be a torus and let $\Delta_n^p$ be the corresponding periodic b.c. Laplace operator. The standard gauge-invariant many body interactions with an interparticle two-body potential $V$ [3, 4] can be decomposed as follows.

$$H_n^{\text{int}} = H_n^{diag} + \delta H_n,$$

where the so called diagonal part $H_n^{diag}$ of $H_n^{\text{int}}$ is explicitly given by

$$H_n^{diag} = \frac{\hat{V}(0)}{2|\Lambda_n|}(N^2 - N) + \frac{1}{2|\Lambda_n|}\sum_k \sum_{k' \neq k} \hat{V}(k - k') n_k n_{k'} \qquad (6)$$

where $N$, $n_k$ are the corresponding number operators, and $\hat{V}$ stands for the Fourier transform of $V$.

It is worthwhile to mention here that a variety of models, such as Huang-Yang-Luttinger model [6], mean field-like models [8], etc., which were intensively analyzed in the past, correspond to certain approximations of $H_n^{diag}$. It seems that the result of [4] on the stability of the Bose-Einstein condensation under the perturbation of the free Bose gas hamiltonian by the diagonal part of $H_n^{\text{int}}$ is one of the main achievements of the (mentioned above) Dublin group's activity and, in a sense, it is a result that is closest to realistic interactions. Having passed throughout the proofs in [4, 7] it becomes clear that the opportunity to express the diagonal part of $H_n^{\text{int}}$ as a bilinear form of the commuting random variables $n_k$ is the basie feature which has enabled such a successful analysis.

The main objective of the present contribution is to apply some, mostly new, commutative tools for analyzing the perturbations of the free Bose matter by thermofield-like interactions (next paragraph), and also by the standard gauge invariant many body interactions. The main result of this paper asserts that non-purity of the free state remains preserved, but under a certain regularity condition, imposed on the limiting thermal state, which can be thought of as exponential type regularity. The last section presents briefly the standard many body bosonic systems in terms of Gibbsian-like perturbations of the introduced Poissone-Wiener random field. It is an interesting problem to be solved here to find a similar description of the ideal Bose gas in the critical regime of couplings.

## II.    GAUSSIAN ANALYSIS: THERMOFIELD PERTURBATIONS

Let $M$ be the Weyl algebra on the Hilbert space $L^2(R^d, dx)$, the generic elements of which we shall denote by $W_f$, $f \in L^2(R^d, dx)$. The following faithful states:

$$\omega_o^\beta(W_f) = \exp\left(-\frac{1}{4}\langle f | \coth\left(\frac{\beta}{2} h_\mu\right) f \rangle_2\right), \qquad h_\mu = -\Delta + \mu, \qquad (7)$$

and (for $d \geq 3$)

$$\omega_{o,cr}^\beta(W_f) = \exp\left(-c|\hat{f}(0)|^2\right) \exp\left(-\frac{1}{4}\langle f | \coth\left(\frac{\beta}{2} h_o\right) f \rangle_2\right), \qquad (8)$$

where $c > 0$, and $\hat{f}$ is the Fourier transform of $f$, are invariant under the evolution

$$\alpha_t f = e^{-ith_\mu} f,$$

and give rise by the GNS construction to the Araki-Wood thermal modules [4]

$$N_{(cr)} \equiv \left(H_{(cr)}^\omega, \Omega_{cr}^\omega, M_{cr}^\omega, \alpha_t^\omega, \beta\right)$$

defined to be free noncritical (resp. critical) modular structures of noninteracting Bose matter. It has been shown in [8, 9] that both $N$ and $N_{cr}$ are stochastically positive and stochastically determined, i.e., there exists process $\xi_\tau$ (resp. $\xi_\tau^{cr}$) with values in the space $S'(R^d)$ ($\equiv$ the space of all tempered distributions) and such that for $f_1,\ldots, f_n$ real, $0 \leq \tau_1 \leq \tau_2 \leq \cdots \leq \tau_n \leq \beta$ the following relationship holds.

$$E e^{i(\xi_{\tau_1}, f_1)} \cdots e^{i(\xi_{\tau_n}, f_n)} = \omega_{o,(cr)}^\beta\left(\alpha_{i\tau_1} W_{f_1} \cdots \alpha_{i\tau_n} W_{f_n}\right) \qquad (9)$$

where on the right hand side stands the corresponding Euclidean time Green function of the thermal structure $N_{(cr)}$. The process $\xi_\tau^{cr}$ is periodic, reflection positive on the circle $K_\beta$ of circumference β, shift invariant and Markovian on $K_\beta$. The above relationships can be realized as the Gaussian measures $d\mu^\beta_{o,(cr)}$ on the space of all continuous loops

$$L^\beta_c(S'(R^d)) = \{\varphi : K_\beta \to S'(R^d) | \varphi \text{ continuous}\},$$

with the covariance explicitly given by

$$\int_{L^\beta_c(S'(R^d))} d\mu^\beta_o(\varphi)\varphi(f \otimes \delta_o)\varphi(g \otimes \delta_\tau)$$

$$= \int dx \int dy\, f(x) \frac{e^{-\tau h_\mu} + e^{-(\beta-\tau)h_\mu}}{1 - e^{-\beta h_\mu}} g(y) \qquad (10)$$

in the noncritical case, and by

$$\int_{L^\beta_c(S'(R^d))} d\mu^\beta_{o,cr}(\varphi)\varphi(f \otimes \delta_o)\varphi(g \otimes \delta_\tau)$$

$$= c\hat{f}(0)\hat{g}(0) + \left\langle f \left| \frac{e^{-\tau h_o} + e^{-(\beta-\tau)h_o}}{1 - e^{-\beta h_o}} g \right. \right\rangle_2 \qquad (11)$$

in the critical one.

It is important to point out here that the canonical procedure of reconstructing the whole noncommutative modular structure of the free Bose matter from the process $\xi_\tau\left(\xi^{cr}_\tau\right)$ is well known [8, 10, 11]. It is the non-ergodicity of the process $\xi^{cr}_\tau$ that reflects the presence of Bose-Einstein condensate (see, e.g., [3, 5] in the critical case: the spontaneous breaking of $U(l)$-gauge invariance). In papers [8, 9, 12] gentle perturbations of the free thermal structure $N_{(cr)}$ have been studied using the idea that the central, affiliated with $M^{\omega''}$ operators act as multiplications in the Gaussian space $L^2\left(L^\beta_c(S'(R^D)), d\mu^\beta_o\right)$. In particular the preservation of the non-ergodicity (in the thermodynamic limit) of the perturbed critical modular structure (thus confirming the stability of Bose-Einstein condensate under such perturbations) has been demonstrated there. Here, we shall report on a recent stability result for perturbations of polynomial types.

Let $a_x$, $a_x^+$ denote annihilation and (respectively) creation operators (in the Araki-Wood module $N_{(cr)}$). The Local Polynomial interaction term is defined by

$$(LP)_\Lambda : \begin{cases} H_\Lambda^{int} = -\lambda \int_\Lambda dx P\left(a_x^\varepsilon + a_x^{+\varepsilon}\right) \\ \lambda \geq 0, \Lambda \subset R^d \text{ is bounded}, P(x) \text{ is a polynomial bounded} \\ \text{from below}, a_x^\varepsilon, a_x^{+\varepsilon} \text{ are properly regularized bosonic} \\ \text{annihilation and creation operators.} \end{cases}$$

The non Local Polynomial interaction term is defined by

$$(nLP)_\Lambda : \begin{cases} H_\Lambda^{int} = -\lambda \int_\Lambda dx\, dy P\left(a_x^\varepsilon + a_x^{+\varepsilon}\right) F(x-y) P\left(a_y^\varepsilon + a_y^{+\varepsilon}\right) \\ \text{where } \lambda \geq 0, \text{ the kernel } F \in L^1(R^d), \\ \text{and it is positive definite.} \end{cases}$$

The perturbed by the local or nonlocal polynomial interactions thermal measure $\mu_{\Lambda,(cr)}^\beta$ is given by the same formula

$$d\mu_{\Lambda,(cr)}^\beta(\varphi) = \left(Z_{\Lambda,(cr)}(\beta)\right)^{-1} \exp\left(\int_0^\beta d\tau \int_\Lambda dx H_\Lambda^{int}(\varphi)(\tau, x)\right) \cdot d\mu_{0,(cr)}^\beta(\varphi), \quad (12)$$

where

$$Z_{\Lambda,(cr)}(\beta) = \int_{S'(K_\beta \times R^d)} d\mu_{\Lambda,(cr)}^\beta(\varphi) \exp \int_0^\beta d\tau \int_\Lambda dx H_\Lambda^{int}(\varphi)(\tau, x). \quad (13)$$

Although one can control the limits $\lim_\Lambda \mu_\Lambda^\beta$ by applying the high temperature cluster expansion [12], the problem of constructing $\lim_\Lambda \mu_{\Lambda,cr}^\beta$ is much more complicated due to the long-range nature of the corresponding potential [13].

Let $\omega_{\Lambda,cr}^\beta$ stand for the state on $M_{cr}''$ obtained by the unitary-like cocycle perturbation of $\omega_{0,cr}^\beta$, where the underlying cocycle $\Gamma_t(\Lambda)$ is given (informally) by

$$\Gamma_t(\Lambda) = \exp it(H^0 + H^{int}_\Lambda)\exp- itH^0 \tag{14}$$

where $H^0$ is the free Bose gas hamiltonian. The main result of [12] is the following one.

**Theorem 2.1.** *If*

$$\sup_\Lambda \int d\mu^\beta_{\Lambda,cr}(\varphi)e^{\varphi(f)} < \infty,$$

*then for sufficiently small $\lambda$ (depending on $\beta$, F,...) there exists a unique limit $\lim_\Lambda \omega_{\Lambda,cr} \equiv \omega^\lambda_{cr}$ (as a weak limit), and a limiting state is faithful and is not a pure state on $M''_{cr}$.*

*The main ideas of the proof.*

**Step 1.** From Araki and Wood's paper [14] (see also [5]) we know that the following decomposition holds:

$$\omega^\beta_{0,cr} = \int d\lambda_0(r,\theta)\omega^{cr}_{r,\theta} \tag{15}$$

where $d\lambda_o(r,\theta)$ is a spectral measure on $[0,\infty) \times K_{2\pi}$, while $\omega^{cr}_{r,\theta}$ are pure quasi-free $\beta$-KMS states on $N^\omega_{cr}$ given by

$$\omega^{cr}_{r,\theta}(W_f) = e^{ic^{1/2}r^{1/2}\cos\theta \cdot \hat{f}(0)} e^{-\frac{1}{2}S^\beta_{\mu=0}(f\otimes f, 0)} \tag{16}$$

where $S^\beta_{\mu=0}$ equals to covariance (10) with $\mu = 0$ setting. Restricting $\omega^{c,r}_{r,\theta}$ to the abelian sector $M^\omega_{cr}$ one can show that, for $f = \bar{f}$, there exist Gaussian probability measures $d\mu^\beta_{r,\theta}$ on $S'(K_\beta \times R^d)$ such that

$$\omega^{cr}_{r,\theta}(W_f) = \int_{S'(K_\beta \times R^d)} d\mu^\beta_{r,\theta}(\varphi)e^{i(\varphi,f)} \tag{17}$$

Thus, we have

$$\omega^\beta_{0,cr}(W_f) = \int d\lambda_0(r,\theta) \int_{S'(K_\beta \times R^d)} d\mu^\beta_{r,\theta}(\varphi)e^{i(\varphi,f)} \tag{18}$$

Then, it follows easily from (15) that

$$\omega_{\Lambda,cr}(W_f) = \int d\lambda_0(r,\theta) \int d\mu_{r,\theta}^{\beta}(\varphi) e^{i(\varphi,f)} \cdot \frac{Z_{\Lambda,cr}^{(r,\theta)}}{Z_{\Lambda,cr}}, \qquad (19)$$

where

$$d\mu_{\Lambda,cr}^{(r,\theta)}(\varphi) = \frac{1}{Z_{\Lambda,cr}^{(r,\theta)}} e^{H_{\Lambda}^{\text{int}}(\varphi)} d\mu_{r,\theta}^{\beta}(\varphi) \qquad (20)$$

with $Z_{\Lambda,cr}^{r,\theta}$ as a normalization factor. (Now and later on we often drop a $\beta$-index for simplicity.)

The Gaussian measures $d\mu_{r,\theta}^{\beta}(\varphi)$ have fast decay of correlations, in contrast to $d\mu_{0,cr}^{\beta}$ and this enables us to prove convergence of the corresponding expansions.

**Proposition 2.2.** *There exists a real $\lambda_0$ (depending on $\beta$,...) such that for $\lambda \in (0, \lambda_0)$ all the corresponding high temperature cluster expansions for $\mu_{\Lambda,cr}^{r,\theta}$ converge uniformly in the parameters $(r, \theta)$. The limiting measures $\mu_{cr}^{r,\theta}(\lambda)$ are ergodic with respect to translations.*

**Step 2.** From the assumption $\sup_{\Lambda} \omega_{\Lambda,cr}(W_f) < \infty$ and uniform convergence $\mu_{\Lambda,cr}^{(r,\theta)} \to \mu_{cr}^{(r,\theta)}(\lambda)$ it follows that there exists the limit (in the sense of measures)

$$\lim_{\Lambda} d\lambda_0(r,\theta) \frac{Z_{\Lambda,cr}^{(r,\theta)}}{Z_{\Lambda,cr}} \equiv d\lambda_{ren}(r,\theta).$$

Moreover, the limiting measure $\lim_{\Lambda} \mu_{\Lambda,cr} \equiv \mu_{cr}^{\lambda}$ is given by

$$\mu_{cr}^{\lambda} = \int d\lambda_{ren}(r,\theta) \mu_{cr}^{(r,\theta)}(\lambda) \qquad (21)$$

The measure $d\lambda_{ren}(r,\theta)$ is not concentrated at one point, which follows from integration by parts formula exactly as in the case of gentle perturbations studied in [8, 9].

**Step 3.** Using a certain reduction formula derived in [13] (in the spirit similar to that presented in [12]), we can prove on the ground of Step 2 that for any

$f \in L^1 \cap L^2(R^d)$, $d \geq 3$, there exists the unique thermodynamic limit

$$\lim_\Lambda \omega_{\Lambda,cr}(W_f) \equiv \omega^\lambda_{cr}(W_f)$$

providing $\lambda$ sufficiently small, and, moreover, the limiting state $\omega^\lambda_{cr}$ can be decomposed into pure states as follows:

$$\omega^\lambda_{cr} = \int d\lambda_{ren}(r,\theta) \omega^{(r,\theta)}_\lambda, \tag{22}$$

where $\omega^{(r,\theta)}_\lambda$ are the appropriate extensions of the measures $\mu^{(r,\theta)}_{cr}(\lambda)$ (viewed as functionals) to the whole Weyl algebra. That would complete the proof.

### III. POISSONIAN ANALYSIS: PERTURBATIONS BY STANDARD GAUGE INVARIANT HAMILTONIANS

In this section we will present a construction of the diagonalizing space for the standard, many body hamiltonian

$$H^{int}_\Lambda = \int_\Lambda a^+(x) a^+(y) V(x-y) a(x) a(y) dx dy \tag{23}$$

where the two-particle potential $V$ obeys standard requirements (see below). Although the underlying construction may be given in the pure Poissonian analysis language [18, 19], we display that construction in terms of Generalized Random Fields (GRF). For this goal, let us denote by $O'_r(p_\beta; \Lambda)$ a (closed) subspace of $D'(R_+ \times \Lambda)$ ($\equiv$ the real Schwartz distributions with support in $R_+ \times \Lambda$. The subspace $O'_r(p_\beta; \Lambda)$ is defined to be the all $\varphi \in D'(R_+ \times \Lambda)$ such that
   (i)  the map $R_+ \ni t \to \varphi(t, \cdot) \in D'(\Lambda)$ is continuous, and
   (ii) there exists $j \in N$ such that $R_+ \ni t \to \varphi(t, \cdot) \in D'(\Lambda)$ is periodic with period $j\beta$.

On the space $D(R_+ \times \Lambda)$ we define the following characteristic functional $\Gamma_\Lambda$:

$$\Gamma_\Lambda(f) \equiv \exp \sum_{j=1}^\infty \frac{z^j}{j} \int_\Lambda dx \int d_\Lambda W^{j\beta,\sigma}_{x|x}(\omega^j) \left( \exp\left( i \int_0^{j\beta} f(\tau, \omega^j(\tau)) d\tau \right) - 1 \right), \tag{24}$$

where $dW^{j\beta,\sigma}_{x|x}$ stands for the $\sigma$-conditioned Brownian bridge measure of length $j\beta$ and the conditioning is given by the classical b.c., i.e., $\sigma \in C(\partial\Lambda)$ and $\sigma \geq 0$ pointwise on $\partial\Lambda$ (see, e.g., [5] for this). By Minlos theorem there exists a measure $dP_\Lambda$ on the space $D'(R_+ \times \Lambda)$ (called a free functional Poisson measure) such that

$$\int_{D'(R_+ \times \Lambda)} dP_\Lambda^\sigma(\varphi) e^{i(\varphi, f)} = \Gamma_\Lambda(f).$$

Actually, it can be shown that the set $O'_r(p_\beta; \Lambda)$ is the carrier set for the measure $dP_\Lambda$. Other elementary properties of $dP_\Lambda$ can be found in [18].

The two-particle potential $V$ in (23) is assumed to obey the following standard ([20]) assumptions:

$c_0$) $V$ is central and $V \in C(R^d \setminus \{0\})$,

$c_1$) $V$ is stable,

$c_2$) there exists $r_0 \geq 0$ such that $\int_{r_0}^\infty |V(r)| dr < \infty$.

**Proposition 3.1.** *Let $V$ fulfill $c_0 - c_2$. Then the following equality is valid:*

$$Z_\Lambda(z, \beta) = Tr_{\Gamma(L^2(\Lambda, dx))} \exp-\beta \left[ d\Gamma\left(-\Delta_\Lambda^\sigma + \mu \mathbf{1}\right) + H_\Lambda^{int} \right] \quad (25)$$

$$= \int_{O'_r(p_\beta; \Lambda)} dP_\Lambda^\sigma(\varphi) \exp-\int_0^\beta d\tau \iint_\Lambda dx\, dy : \varphi(x) V(x-y) \varphi(y):$$

where *(informally)*

$$:\varphi(x) V(x-y) \varphi(y) := \varphi(x) V(x-y) \varphi(y) - V(0) \# \{\varphi\},$$

and $\# \{\varphi\}$ is a „loop number" functional defined (for $dP_\Lambda^\sigma$ a.e. $\varphi$) in [15,16]. *Proof. See* [17,18]

Thus, in a certain sense, the space $L^2(dP_\Lambda^\sigma)$ plays the role of a diagonalizing space for $H_\Lambda^{int}$. The Gibbsian type perturbation $dG_\Lambda^\sigma$ of $dP_\Lambda^\sigma$ is defined by

$$dG_\Lambda^\sigma(\varphi) = Z_\Lambda(z, \beta)^{-1} \exp-\int_0^\beta d\tau \iint_\Lambda dx\, dy : \varphi(x) V(x-y) \varphi(y): dP_\Lambda^\sigma. \quad (26)$$

**Theorem 3.2.** Let $V$ satisfy $c_0 - c_2$. Then, for sufficiently small $z$ (depending on $\beta, V,...$), a unique thermodynamic limit

$$\lim_{\Lambda \uparrow R^d} \int dG_\Lambda^\sigma(\varphi) \prod_{i=1}^n (\varphi, f_i) \equiv \int dG(\varphi) \prod_{i=1}^n (\varphi, f_i) \qquad (27)$$

exists for any $n \in N$, $(\eta^{j_1},...\eta^{j_n})$, and the limits on l.b.s of the above formula uniquely determine the measure $dG(\varphi)$ in the integral on r.h.s. Moreover, the limiting measure $dG$ does not depend on the particular choice of $\sigma$.

*The main ideas of the proof.*

**Step l.** The following formula for integration by parts holds:

$$\int dP_\Lambda^\sigma(\varphi) :\langle \varphi, f \rangle F(\varphi):_P G(\varphi) = \sum_{j=1}^\infty \frac{z^j}{j} \int_\Lambda dx \int d_\Lambda W_{x|x}^{j\beta\sigma}(\omega^j) \int_0^{j\beta} d\tau\, f(\tau, \omega^j(\tau))$$

$$\cdot \int dP_\Lambda^\sigma(\varphi): F(\varphi):_P G(\varphi + \delta(\cdot - \omega^j(\tau))), \qquad (28)$$

for any cylindrical, $L^1(dP_\Lambda)$ - functionals $F$ and $G$, where $::_P$ denotes the Poissonian normal ordering (see, e.g., [17, 18]).

**Step 2.** Applying Step l, we get the relationship

$$\int dG_\Lambda^\sigma(\varphi): \prod_{i=1}^n (\varphi, f_i):_P = \sum_{j_1,...,j_n \geq 1} \frac{z^{j_1+...+j_n}}{j_1 \cdots j_n} \prod_{\alpha=1}^n \int_\Lambda dx_\alpha \int d_\Lambda W_{x_\alpha|x_\alpha}^{j_\alpha\beta,\sigma}(\omega^{j_\alpha})$$

$$\cdot \int_0^{j_\alpha\beta} f(\tau, \omega^{j_\alpha}(\tau)) d\tau \cdot \overset{\bullet}{\sigma}_\Lambda(\omega^{j_1},...,\omega^{j_n}) \qquad (29)$$

where

$$\overset{\bullet}{\sigma}_\Lambda(\omega^{j_1},...,\omega^{j_n}) = e^{-\varepsilon_V^\beta(\omega^{j_1},...,\omega^{j_n})} \int dG_\Lambda^\sigma(\varphi) \exp{-\varepsilon_V^\beta(\omega^{j_1},...,\omega^{j_n}|\varphi)}, \qquad (30)$$

$$\varepsilon_V^\beta(\omega^{j_1},...,\omega^{j_n}) = \sum_{1 \leq k < l \leq n} \int_0^\beta d\tau \sum_{\alpha_k=0}^{j_k-1} \sum_{\alpha l=0}^{j_l-1} V(\omega^{j_k}(\tau+\alpha_k\beta); \omega^{j_l}(\tau+\alpha_l\beta)) \qquad (31)$$

$$\varepsilon_V^\beta\left(\omega^{j_1},...,\omega^{j_n}|\varphi\right) = \sum_{k=1}^{n} \varepsilon_V^\beta\left(\omega^{j_k};\varphi\right), \qquad (32)$$

and $\varepsilon_V^\beta\left(\omega^{j_k};\varphi\right)$ is defined ($dP_\Lambda^\sigma$ a.e.) in [15,16]. Now, let $P_\Lambda^\sigma(j_1,...,j_n)$ be the space consisting of $n$-tuples of paths $\left(\eta^{j_1},...,\eta^{j_n}\right)$ where

$$\eta^{j_k} \in C\left([0, j_k\beta] \to \overline{\Lambda}\right).$$

The extensions of functionals $\varepsilon_V^\beta\left(\cdot|\varphi\right)$ to the spaces $P_\Lambda^\sigma(j_1,...,j_n)$ and the corresponding extensions of functionals $\sigma_\Lambda^\bullet$ given by (30) will be denoted by the same symbols.

**Step 3.** Let $\sigma_\Lambda^\bullet$ be the extensions of the functionals (30). Then, the standard $\sigma$-conditioned (see, e.g., [5]) reduced density matrices $\rho_\Lambda^\bullet$ are given by their kernels

$$\rho_\Lambda^\bullet(x_1,...,x_n|y_1,...,y_n) =$$

$$z^n \sum_{\pi \in S_n} \sum_{j_1=1}^{\infty} \cdots \sum_{j_n=1}^{\infty} \int d_\Lambda W_{x_1|\pi(y_1)}^{j_1\beta,\sigma}\left(\omega^{j_1}\right) \cdots \int d_\Lambda W_{x_n|\pi(y_n)}^{j_n\beta,\sigma}\left(\omega^{j_n}\right)$$

$$\cdot \frac{z^{j_1+\cdots+j_n}}{j_1 \cdots j_n} \sigma_\Lambda^\bullet\left(\omega^{j_1},...,\omega^{j_n}\right)$$

Therefore, the (extended to deal with the case of arbitrary classical b.c.) classical analysis of Ginibre [21, 22] can be applied to control the limits $\lim_\Lambda \sigma_\Lambda^\bullet(\cdot)$ rigorously.

**Step 4.** Comparison of the corresponding Kirkwood-Salsburg resolvent expansions leads to a demonstration of $\sigma$-independence of the limiting Gibbs measures $dG$.

## IV.    REFERENCES

[1] A. Einstein, Sitzungsberichte der Preussischen Akademie der Wissenschaften l (1925), 3.
[2] R. M. Ziff, G.E., Uhlembeck, M. Kac, Phys. Rep. **32 C** no. 4 (1977).
[3] M. Van den Berg, J.T. Lewis, J.V. Pule, Helv. Phys. Acta **59** (1986), 1271.
[4] T.C. Dorlas, J.T. Lewis, J.V. Pule, Commun. Math. Phys. **156** (1993), 37.
[5] O. Bratelli, D.W. Robinson, *Operator Algebras and Quantum Statistical Mechanics,*

*II,* Springer, Berlin, 1981.
[6] K. Huang, C.N. Yang, J.A. Luttinger, Phys. Rev. **105** (1957), 776.
[7] T.C. Dorlas, J.T. Lewis, J.V. Pule, Helv. Phys. Acta **64** (1991), 1200.
[8] R. Gielerak, R. Olkiewicz, J. Stat. Phys. **80** (1995), 875.
[9] ___, J. Math. Phys. **37** (1996), 1268.
[10] R. Gielerak, L. Jakóbczyk, R. Olkiewicz, J. Math. Phys. **39** (1998), 6291.
[11] ___, J. Math. Phys. **35** (1994), 6291.
[12] R. Gielerak, *Some models of nonideal Bose gas with Bose-Einstein condensate,* Mathematical Physics and Stochastic Analysis. Essays in Honour of Ludwig Streit., Worid Scientific, 2000, p, 215.
[13] R. Gielerak, J. Damek, *Thermofield, polynomial perturbations of the ideal Bose gas,* in preparations.
[14] H. Araki, E.J. Wood, J. Math. Phys. **4** (1963), 637.
[15] R. Gielerak, R. Rebenko, J. Math. Phys. **37** (1996), 3354.
[16] S. Albeverio, Yu. Kondratiev, M. Rökner, J. Funct. Anal. (1998).
[17] R. Gielerak, R, Rebenko, Oper. Theory: Advances and Applications 70 (1994), 219.
[18] R. Gielerak, *Gibbsian approach to quantum many body problems,* paper(s) in preparations.
[19] E.W. Lytvynov, A.L. Rebenko, G.V. Schepanyuk, Rep. Math. Phys. **37** (1996), 157.
[20] D. Ruelle, *Statistical Mechanics. Rigorous Results.,* Benjamin, New York, 1969.
[21] J. Ginibre, J. Math. Phys. **6** (1965), 238.
[22] J. Ginibre, *Some applications of functional integration in Statistical mechanics,* Statistical Mechanics and Field Theory (C. de Witt, R. Stora, eds.), Gordon and Breach, New York, 1971.